\documentclass
[aps,prd,nofootinbib,
superscriptaddress,preprintnumbers,
twocolumn
]{revtex4-1}

\usepackage{hyperref}
\usepackage{color}
\usepackage{graphicx,subfigure}
\usepackage{siunitx}
\usepackage{amsmath,amssymb,amsfonts}
\usepackage{bbold}
\usepackage{bm}
\usepackage{hyperref}
\usepackage{ulem}
\usepackage{lineno}
\usepackage{soul}

\graphicspath{{./plot/}}

\def \be {\begin{equation}}
\def \ee {\end{equation}}
\def \bes {\begin{subequations}}
\def \ees {\end{subequations}}

\newcommand{\beq}{\begin{eqnarray}}
\newcommand{\eeq}{\end{eqnarray}}

\def \pd {\partial}
\newcommand{\Eq}[1]{Eq.~\eqref{#1}}
\newcommand{\Fig}[1]{Fig.~\ref{#1}}

\newcommand{\sH}{\mathcal{H}}
\newcommand{\sL}{\mathcal{L}}

\def \f {f \left(p_z,p_\perp;\tau \right)}
\def \Fe{F_{\e} \left(\cos\theta;\tau \right)}

\def \<{\langle}
\def \>{\rangle}
\def \+{\dagger}
\def \({\left(}
\def \){\right)}
\def \[{\left[}
\def \]{\right]}
\def \a {\alpha}
\def \b {\beta}

\def \d {\delta}
\def \e {\epsilon}

\def \l {\lambda}

\def \t {\tau}

\def\tE {{\cal E}}
\def \name{``adiabatic hydrodynamization''\,}
\def \nameshort {AH}

\def \keyf {g}

\def \tnew {y}

\def\tAtt {\tau_{{\rm Redu}}}
\def\tHydro {\tau_{\rm Hydro}}

\def\tColl{\tau_{{\rm C}}}

\newcommand\sect[1]{\noindent \textbf{#1.}---}

\begin{document}

\preprint{MIT-CTP/5141}

\title{
Adiabatic hydrodynamization in rapidly-expanding quark--gluon plasma
}

\author{Jasmine Brewer}
\affiliation{Center for Theoretical Physics, Massachusetts Institute of Technology, Cambridge, Massachusetts 02139, USA }

\author{Li Yan}
\affiliation{Key Laboratory of Nuclear Physics and Ion-Beam Application (MOE) \& Institute of Modern Physics,
Fudan University, 220 Handan Road, 200433, Yangpu District, Shanghai, China}

\author{Yi Yin}
\affiliation{Center for Theoretical Physics, Massachusetts Institute of Technology, Cambridge, Massachusetts 02139, USA }
\affiliation{Quark Matter Research Center, Institute of Modern Physics, 
Chinese Academy of Sciences,
 Lanzhou, Gansu, 073000, China }
 \affiliation{
 University of Chinese Academy of Sciences,
 Beijing, 100049, China
 }

\date{\today}

\begin{abstract}

We propose a new scenario characterizing the transition of the quark–gluon plasma (QGP) produced in heavy-ion collisions from a highly non-equilibrium state at early times toward a fluid described by hydrodynamics at late times. We develop an analogy to the evolution of a quantum mechanical system that is governed by the instantaneous ground states. In the simplest case, these slow modes are “pre-hydrodynamic” in the sense that they are initially distinct from, but evolve continuously into, hydrodynamic modes. For a class of collision integrals, the pre-hydrodynamic mode represents the angular distribution (in momentum space) of those gluons that carry most of the energy. We illustrate this scenario using a kinetic description of weakly-coupled Bjorken expanding plasma. Rapid longitudinal expansion drives a reduction in the degrees of freedom at early times. In the relaxation time approximation for the collision integral, we show quantitatively that the full kinetic theory evolution is dominated by the pre-hydrodynamic mode. We elaborate on the criterion for the dominance of pre-hydrodynamic slow modes and speculate that adiabatic hydrodynamization may describe the pre-equilibrium behavior of the QGP produced in heavy-ion collisions.
\end{abstract}

\maketitle

\section{Introduction}

Hydrodynamics describes the real-time dynamics of a broad class of interacting many-body systems in the long time and long wavelength limit.
In this limit, most degrees of freedom become irrelevant since they relax on short time scales. 
The surviving slow dynamical variables, 
or ``hydrodynamic modes'', are those associated with conserved densities such as the energy density. 
Hydrodynamic modelling has seen remarkable success at describing varied and non-trivial results of heavy-ion collision experiments (see Ref.~\cite{Heinz:2013wva} for a concise review). This in turn raises the important question of how the system approaches a state dominated by hydrodynamic modes, namely how ``hydrodynamization'' occurs in the aftermath of a heavy-ion collision (cf. \cite{Romatschke:2017ejr,Heller:2016gbp,Florkowski:2017olj} for a recent review).

In this letter, 
we theorize a new scenario for the process of hydrodynamization with the following defining attribute:
\textit{
during the interval $\tAtt < \tau < \tHydro$, the bulk evolution is governed by a set of slow modes that are ``pre-hydrodynamic'' in the sense that they are distinct from hydrodynamic modes but evolve gradually into them around the time $\tHydro$.}
As a premise of this picture, we assume the emergence of a time scale $\tAtt<\tHydro$ around which the degrees of freedom required to describe the bulk properties of the system are reduced (see more below).

The pre-hydrodynamic modes in the preceding scenario are the modes with the slowest rate of change at each instant in the pre-hydrodynamic evolution, under the assumption that they remain gapped from faster modes.
They are closely analogous to the instantaneous ground states of a time-dependent Hamiltonian in quantum mechanics. 
Since near thermal equilibrium the hydrodynamic modes are the slowest modes, 
the pre-hydrodynamic modes are a natural off-equilibrium generalization of the hydrodynamic modes.
If a time-dependent and gapped quantum-mechanical system is prepared in its ground state, it will remain in the instantaneous ground state under adiabatic evolution of the Hamiltonian. 
We will thus refer to situations where the approach to hydrodynamics is governed first by the evolution of pre-hydrodynamic modes as \name (\nameshort).

We will illustrate AH in a kinetic description of weakly-coupled Bjorken-expanding plasma. 
We explicitly identify the pre-hydrodynamic modes as the instantaneous ground state modes of a non-Hermitian matrix describing the evolution of bulk quantities from the kinetic equation with a class of collision integrals.
Physically, these modes represent the angular distribution in momentum space of the gluons that carry most of the energy of the system.  
We then demonstrate the emergence of $\tAtt$ induced by the fast longitudinal expansion and show that $\tAtt$ is parametrically smaller than $\tColl$.
This is due to the separation of scales between the initial time $\tau_I$ when the kinetic description becomes applicable and the typical collision time $\tColl$ for QGP in the weak-coupling regime. 
Because of the hierarchy $\tAtt\ll \tColl\ll\tHydro$, 
the pre-history of hydrodynamics is (almost) the history of pre-hydrodynamic modes within AH.

An important implication of \nameshort\ is that
the macroscopic properties of the medium during the pre-hydrodynamic stage are insensitive to both the initial conditions and the details of the expansion history, and instead are determined predominantly by the features of the pre-hydrodynamic modes.
In particular, 
the most important quantity characterizing the bulk evolution of a plasma undergoing Bjorken expansion is the percentage rate of change of the energy density
\begin{eqnarray}
\label{key-fun-def}
\keyf(\tnew)&\equiv& - \pd_{y}\log\e\, , 
\end{eqnarray}
where $\tnew\equiv \log(\tau/\tau_{I}) $ plays the role of a time variable. 
We shall show that for Bjorken expansion $\keyf(\tnew)$ is related to the eigenvalue $\tE_{0}(y)$ of the pre-hydrodynamic mode if \nameshort\ applies, 
namely
\begin{eqnarray}
\label{main-result}
\keyf(y)&\approx& \tE_0(y)\, .
\end{eqnarray} 
We consider the extensively-studied relaxation time approximation (RTA) of the kinetic equation~\cite{Florkowski:2013lya,Romatschke:2015gic,Heller:2016rtz,Denicol:2016bjh,Denicol:2017lxn,Heller:2018qvh} and confirm quantitatively that \Eq{main-result} holds, demonstrating that
hydrodynamization in this model is an example of \nameshort.

Because of the expansion, 
the criterion for the dominance of pre-hydrodynamic modes is not that the excited states have decayed, but rather that transitions to the excited states are suppressed. 
In the absence of better terminology, 
throughout this manuscript we will use ``adiabaticity" as a synonym for the suppression of these transitions. 
This is consistent with the modern use of this terminology in quantum mechanics (c.f.~Ref.~\cite{APT}). 
We will show that the regime where this generalized notion of adiabaticity may not apply is parametrically narrow according to the scenario of bottom-up thermalization for weakly-coupled QGP~\cite{Baier:2000sb}. 
Although our analysis relies on the smallness of $\a_{s}$, 
we hope that many qualitative features of AH may nonetheless be present in the QGP created in heavy-ion collisions.

There are extensive studies on the formulation of far-from-equilibrium hydrodynamics to describe 
the pre-hydrodynamic stage of heavy-ion collisions~\cite{Romatschke:2017vte, Heller:2015dha, Strickland:2017kux,Jaiswal:2019cju,Denicol:2017lxn,Behtash:2018moe,Blaizot:2019scw}.
The key premise of this paradigm is that hydrodynamic modes dominate the bulk evolution, and consequently that hydrodynamics is applicable, even when the system is far from equilibrium~\cite{Romatschke:2017vte}.
The difference between this paradigm and \nameshort\ is that the dominant slow modes for systems undergoing \nameshort\ are pre-hydrodynamic mode(s), which can be qualitatively distinct from hydrodynamic modes.
The slow modes of a system generally depend on the state of the medium under consideration, and therefore it is unsurprising that the slow modes in a far-from-equilibrium system are generically different from the hydrodynamic modes. An illustrative example is the emergence of apparent slow degrees of freedom in the approach to the far-from-equilibrium scaling regime in QCD kinetic theory~\cite{Mazeliauskas:2018yef}. Another example is the low-energy collective excitations in a normal Fermi liquid. 
When the frequency of the distribution function variation, an analog of expansion rate, 
is much larger than the collision rate and thermal equilibrium is not established in each volume element, the slow modes are zero sound modes which have different physical characteristics than ordinary sound \cite{landau1980statistical}.
We show that Bjorken-expanding RTA kinetic theory is an example where the pre-hydrodynamic and hydrodynamic modes are qualitatively different, and the bulk evolution of the system is dominated by the pre-hydrodynamic mode.

In the modern view, 
hydrodynamics is a macroscopic effective theory in which hydrodynamic modes are the relevant low energy degrees of freedom.
In cases where the relevant degrees of freedom are actually pre-hydrodynamic modes, there is no guarantee that hydrodynamics or its simple generalizations will describe the system, just as hydrodynamics does not describe the physics of zero sound. This is not in contradiction to the recent result~\cite{Jaiswal:2019cju,Chattopadhyay:2018apf} that some non-trivial generalizations of hydrodynamics like an improved version of Israel-Stewart theory~\cite{Denicol:2012cn} and anisotropic and third-order hydrodynamics describe the bulk evolution of several simplified kinetic theory models even beginning at $\tAtt$. Rather, since these models include significant contributions from non-hydrodynamic modes, we emphasize that this observation alone does not imply that hydrodynamic modes dominate the evolution.
It is worth exploring the applicability of these models in more general settings, 
however we hope that the identification of pre-hydrodynamic modes as a relevant slow degree of freedom may motivate the future construction of an effective theory of ``pre-hydrodynamics".

\section{Pre-hydrodynamic modes}

\subsection{Identification of pre-hydrodynamic mode(s)}

We consider a Bjorken-expanding medium of massless particles 
described by the kinetic equation 
\begin{equation}
\label{eq:kinetic}
\frac{\partial}{\partial\tau}\, \f =
-\frac{p_z}{\tau} \frac{\partial}{\partial p_z} \f
- {\hat C}[f]\,,
\end{equation}
where $\f$ is the single particle distribution, $p_{\perp}$ and $p_{z}$ are the transverse and longitudinal momentum,
and  ${\hat C}$ is the collision integral.
Because of the symmetry, the only relevant hydrodynamic mode is the energy density $\e$.  
To more directly study the evolution of the energy density, we will focus on the momentum-weighted distribution function
\begin{equation}
\label{eqn:Fe-def}
\Fe
\equiv \frac{1}{2\pi^2}\int^{\infty}_{0}\, dp\, p^{3}\, 
\f\, ,
\end{equation}
where $p=\sqrt{p^{2}_{\perp}+p^{2}_{z}}$ and $\theta=\tan^{-1}\(p_{z}/p_{\perp}\)$.
Because the angular integration of $\Fe$ is the energy density, $\Fe$ describes the angular distribution of the particles that carry most of the energy. 
The $p^3$-weighted moment of \Eq{eq:kinetic} gives the evolution equation for $F_\e$:
\begin{align}
 	\label{Fe-evol}
 	\tau \frac{\pd}{\pd \t} &\Fe = 
 	- \left[ -4 \cos^2 \theta + \sin^2 \theta \cos \theta \frac{\pd}{\pd \cos \theta} \right]  \nonumber \\&
 	\times \Fe - \frac{\tau}{2 \pi^2} \int_0^\infty dp \, p^3 \, \hat{C}[f].
\end{align}

Following Ref.~\cite{Blaizot:2017ucy}, we assume that $\f$ is symmetric under $p_{z}\to - p_{z}$ 
and expand $\Fe$ in a basis of the Legendre polynomials $P_{2n}$:
\begin{equation}
\label{F-expand}
    \Fe = \e(\tau) +\sum_{n=1} \frac{4n+1}{2} \sL_n(\tau) P_{2n}(\cos\theta)\, . 
\end{equation}
\Eq{F-expand} maps $\Fe$ to an infinite-dimensional vector $\psi=\(\e, \sL_{1},\sL_{2}, \ldots,\)$. 
We therefore have the correspondence
\begin{eqnarray}
\label{Fe-mapping}
F_{\e}\(\cos\theta\)\leftrightarrow \psi=\(\e, \sL_{1},\sL_{2}, \ldots,\)\, .
\end{eqnarray}
Note $p_{L}=\frac{1}{3}\(\e+2\sL_{1}\)$.
Since $F_\e$ will become isotropic and approach $\e$ at late times while $\sL_{n>0}$ are suppressed by $\tau_{C}/\tau\ll 1$, 
we expect that in the hydrodynamic regime
\begin{eqnarray}
\psi \to \phi^{H}_0+{\cal O}\left(\frac{\tau_{C}}{\tau}\right)\, ,
\qquad
\phi^{H}_{0}\equiv\left(\e,0,0,\ldots\right)\, . 
\end{eqnarray}
The problem of hydrodynamization is therefore reduced to studying how $\psi$ approaches $\phi^{H}_0$.

In the following discussion, we shall limit ourselves to the class of collision integrals for which 
Eq.~\eqref{Fe-evol} can be recast into the form
\begin{equation}
\label{psi-eqn}
\pd_{y}\psi=-\,\sH(y)\, \psi\, ,
\end{equation}
where $\sH$ is a non-Hermitian matrix and $y=\log(\tau/\tau_{I})$ as we introduced earlier.
This is trivially satisfied for any collision integral that
is linear in $\Fe$. The discussion here can also be applied to nonlinear collision kernels for which $\sH$ can be expressed as a functional of the distribution function. We largely postpone this discussion to follow-up work \cite{bruno}, though we will briefly discuss the simple case that the relaxation time depends on time through the energy density.
\Eq{psi-eqn} has the structure of the time-dependent Schr\"odinger equation in quantum mechanics.
The explicit expression for the matrix $\sH$ in RTA kinetic theory will be given in the subsequent section.

Throughout this work, we will study the instantaneous eigenmodes $\phi_n(\tnew)$ of $\sH(\tnew)$.
For clarity we will order them by the real part of their corresponding eigenvalues, e.g.
${\rm Re}\tE_{0}< {\rm Re}\tE_{1}\leq\ldots$.
Of particular importance is the ground state mode $\phi_0(y)$, which has the lowest damping rate of all of the eigenmodes.
In the hydrodynamic limit $\tau\geq \tHydro$, the ground state will be $\phi^{H}_{0}$ since the conserved densities are the zero-modes of any collision kernel. 
At times $\tau < \tHydro$, we identify $\phi_0(\tnew)$ as the pre-hydrodynamic mode that evolves into the hydrodynamic mode $\phi^{H}_0$ in the hydrodynamic limit.

\subsection{Emergent dominance of pre-hydrodynamic modes at $\t\ll \tColl$}

To illustrate the reduction in the degrees of freedom at early times, we study the behavior of $\psi$ for $\tau\ll \tColl$.
In this case, $\psi$ is determined by $\pd_y\psi=-\sH_{F}\,\psi$ where $\sH_F$ is obtained from \Eq{Fe-evol} by neglecting the collision integral (the explicit expression can be obtained from the $\t \rightarrow 0$ limit of \Eq{psi-component}).
Then, we expand $\psi$ in eigenstates of $\sH_F$ as 
$\psi(\tau)=\sum_{n=0}\, \b_{n}(\tau) \phi^{F}_{n}$. 
It is easy to show that $\b_{n}(\tau)=\b_{n}(\tau_{I})\, \exp(-\tE_{n}\tnew)$ for all $n$. 
Therefore contributions from the ``excited'' modes $\phi^F_{n>0}$ become unimportant after some emergent time scale
\begin{eqnarray}
\tAtt \gtrsim \tau_{I}\,
\Bigg |
\(\frac{\b_n(\tau_I)}{\b_{0}(\tau_I)}\)^{1/(\tE^{F}_{n>0}-\tE^{F}_{0})}
\Bigg|\, . 
\end{eqnarray}
The bulk evolution of the system around $\tAtt$ is then dominated by the ground state mode $\phi_0^F$.
Related observations have also been made in Refs.~\cite{Jaiswal:2019cju,Kurkela:2019set}.

For the description of heavy-ion collisions in the framework of perturbative QCD,
$\tau_I$ is of the order of $Q_s^{-1}$, 
where $Q_s\gg \Lambda_{\rm QCD}$ is the saturation scale (c.f.~Refs.~\cite{Blaizot:1987nc,McLerran:1993ni,McLerran:1994vd,JalilianMarian:1996xn}). 
Meanwhile, 
a parametric estimate of $\tColl$ can be deduced from the collision integral, $\tColl Q_s\sim \alpha_s^{-x}$ with exponent $x>0$ (c.f.~Ref.~\cite{Baier:2000sb}).
This hierarchy guarantees the existence of a time scale $\tAtt$ that is parametrically smaller than $\tHydro \geq \tColl$.

To appreciate the physics underlying the dominance of $\phi^{F}_0$ around $\tAtt$, 
we compare the explicit expression $\phi^{F}_{0}=\e\(1, P_{2}(0),P_{4}(0)\,\ldots\)$~\cite{Blaizot:2019scw} with the definition in \Eq{F-expand}. 
It is then transparent that $\phi^{F}_{0}$ corresponds to an angular distribution function $\Fe$ that is sharply peaked at $\theta=\pi/2$.
For such a distribution, typical values of $p_z$ are much smaller than those of $p_\perp$,
meaning the longitudinal expansion drives arbitrary initial conditions to a highly anisotropic distribution in momentum space.

The analysis above shows that the longitudinal expansion together with the intrinsic hierarchy $\tau_I\ll \tColl$ in weakly coupled QCD prepares the system in the instantaneous ground state $\phi^F_0$.
Since $\phi^F_0$ depends on $\sH_{F}$ but not on the initial conditions, 
the bulk evolution around $\tAtt$ becomes insensitive to the details of the initial conditions. 
The latter has been observed in previous studies of kinetic theory
~\cite{Romatschke:2017vte,Kurkela:2019kip, Blaizot:2019scw}, 
though its connection to the dominance of the mode $\phi^F_{0}$ has not been elucidated before.

\subsection{Implications of the dominance of  pre-hydrodynamic modes}

We now explore the implications of the adiabatic evolution of $\sH(y)$ after $\tAtt$. 
We begin by expanding $\psi$ in terms of the instantaneous eigenmodes $\phi_{n}(y)$ of $\sH(y)$,
$\psi(\tnew)
= 
\sum_{n=0}\, \a_{n}(\tnew)\, \phi_{n}\(\tnew\)$. 
While in general $\a_{n>0}$ can be the same order of magnitude as $\a_{0}$, under adiabatic evolution $|\a_0|\gg |\a_{n>0}|$ and consequently
\begin{eqnarray}
\label{AH}
\psi(y) \sim \phi_0(y)\, . 
\end{eqnarray}
\Eq{AH} can be viewed as the definition of adiabatic hydrodynamization.

We emphasize that this dominance of the pre-hydrodynamic mode $\phi_0$ indicates that the bulk properties of the pre-equilibrium medium can be related to this mode and its eigenvalue.
For example,
let us focus on the percentage rate of change of the energy density in \Eq{key-fun-def}.
Since the zeroth component of $\psi$ is $\e$, it follows from Eqns.~\eqref{psi-eqn} and \eqref{AH} that $-\e \keyf(y)$ is given by the zeroth component of $\sH \psi$, i.e. Eq.~\eqref{main-result},  
even though $\keyf(y)$ in general can depend on all modes $\phi_n(\tnew)$.
This non-trivial relation is a consequence of the adiabatic evolution.
In the next section we will quantitatively test this result in the relaxation time approximation to determine the extent to which \nameshort\ applies.

\begin{figure*}
\subfigure{
\label{fig:g}
  \includegraphics[width=.45\textwidth]
  {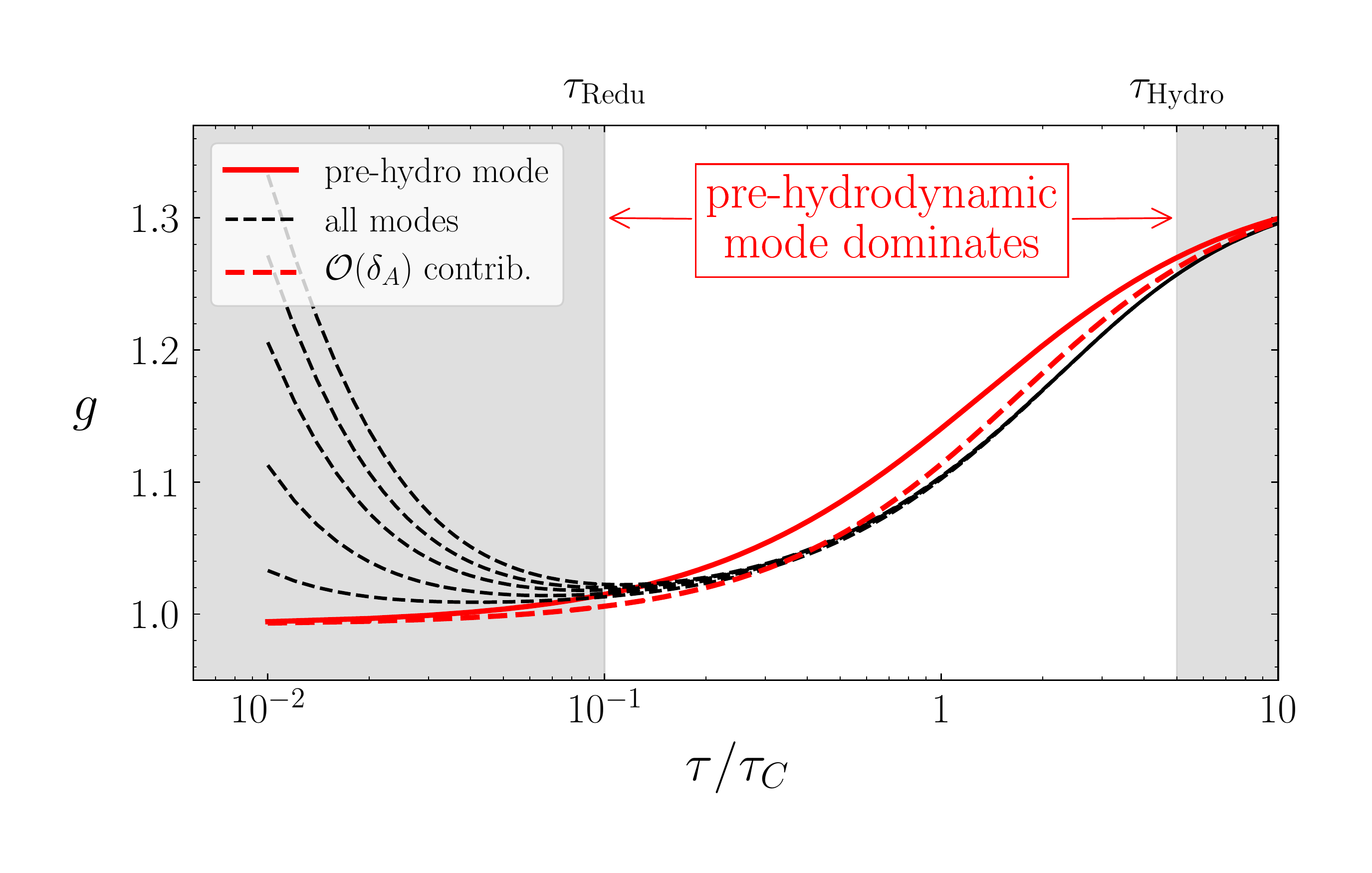}
}
\subfigure{
\label{fig:diff}
  \includegraphics[width=.45\textwidth]{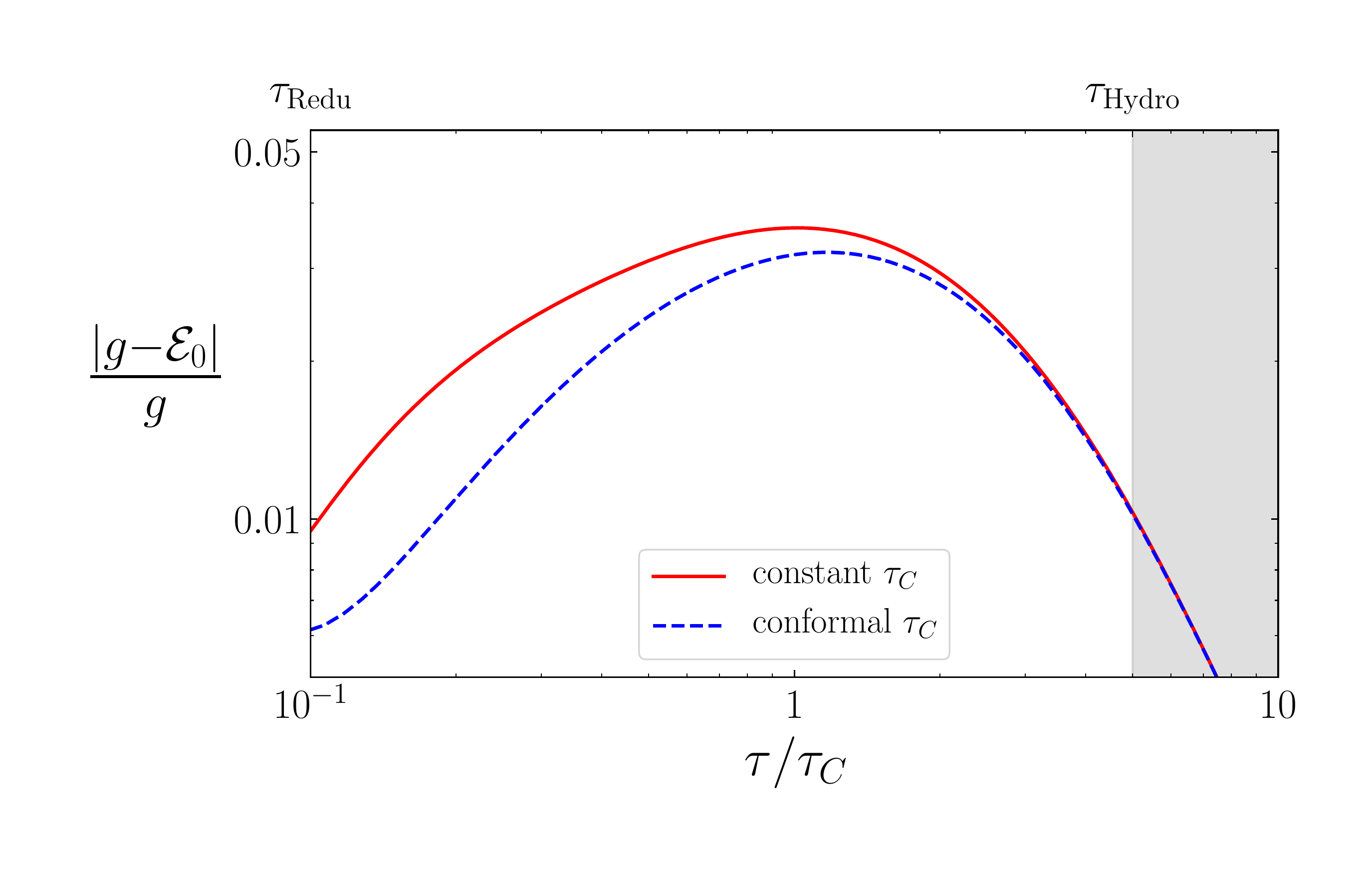}
}
\caption{
We demonstrate the dominance of the pre-hydrodynamic mode, and hence adiabatic hydrodynamization, for a Bjorken-expanding plasma in the relaxation time approximation.
(a) 
The red solid curve shows $\tE_{0}(\l)$, which is the contribution to the percentage rate of change of the energy density coming from the pre-hydrodynamic mode (c.f.~\Eq{main-result}). 
Black curves are $g(\l)$ obtained from solving \Eq{eq:kinetic} with constant $\tColl$.
After $\tAtt$, they collapse onto the RTA attractor as obtained in Refs.~\cite{Heller:2016rtz,Romatschke:2017vte,Heller:2018qvh}.
The dashed red curve is $g(\l)$ obtained by including leading-order contributions from the excited states.
The left and right shaded regions indicate $\tau\leq \tAtt$ and $\tau\geq \tHydro$, respectively. 
(b) 
The fractional difference between $\tE_{0}$ and $g$, 
which measures the relative importance of contributions from the ground state (pre-hydrodynamic) and excited modes.
Red and dashed blue curves show this difference for representative initial conditions with constant and conformal $\tColl$, respectively.
The fact that this quantity is small 
indicates the dominance of the pre-hydrodynamic mode (i.e. adiabaticity) during the interval $\tAtt\leq \tau < \tHydro$. 
}
\label{fig:fig1}
\end{figure*}

\section{RTA as an example of adiabatic hydrodynamization}

The collision integral under the relaxation-time approximation (RTA) is
\begin{equation}
\label{RTA-C}
    \hat{C}[f] = \frac{\f - f_{\rm eq}(p/T)}{\tColl(y)}\, ,
\end{equation}
where $\tColl$ is a function of $y$. 
Substituting Eqns.~\eqref{RTA-C} and \eqref{F-expand} into \Eq{Fe-evol} gives~\cite{Blaizot:2017ucy}
\begin{equation}
\label{psi-component}
     \pd_\tnew \sL_n = - \left[ a_n \sL_n + b_n \sL_{n-1} + c_n \sL_{n+1} \right]- \l (1-\d_{n0}) \sL_n,
\end{equation}
where $\l \equiv \tau/\tColl$. 
Explicit expressions for $a_{n},b_{n}$, and $c_{n}$ are given in Ref.~\cite{Blaizot:2017ucy},
for example $(a_0, b_1,c_0)=(4/3,8/15,2/3 )$.
From \Eq{psi-component}, the evolution of $\psi$ has the form \Eq{psi-eqn} with
\begin{equation}
\label{H-RTA}
\sH(\tnew)=\sH_{{\rm RTA}}(\l)\equiv \sH_{F}+\l\(\tnew\) \sH_{1}\, , 
\end{equation}
where the elements of $\sH_{F},\sH_{1}$ can be read from \Eq{psi-component}.

From $\sH_{{\rm RTA}}(\l)$ we compute the pre-hydrodynamic mode $\phi_{0}(\l)$ and its energy $\tE_{0}\(\l\)$ for each $\l$.
We note that the minimum gap $\Delta \tE_{\min}(\l)\equiv {\rm Re}(\tE_{1}(\l)-\tE_{0}(\l))$ is order one for $\l\ll 1$ and becomes linear in $\l$ for $\l\gg 1$
\footnote{
If we were using $\tau$ instead of $y$ as our temporal variable, 
the minimum gap is of the order $1/\tau$ for $\l\ll 1$ and of the order $1/\tColl$ for $\l \gg 1$. 
The latter agrees with Ref.~\cite{Romatschke:2015gic} although one has to keep in mind that the minimum gap also evolves in time. 
}.
For all values of $\l$, $\phi_{0}(\l)$ is gapped from the excited modes. 
It is easy to check that $\phi^{H}_{0}$ is the ground state of $\sH_{1}$ but not that of $\sH_{{\rm RTA}}$. 
Since $\sH_{{\rm RTA}}$ evolves in time, 
the components of $\phi_{0}(y)$ are different from those of $\phi^{H}_{0}$ for any finite $y$, 
exemplifying the distinction between the pre-hydrodynamic and hydrodynamic modes.

The solid red curve in \Fig{fig:g} shows $\tE_{0}\(\l\)$, which is the contribution to $g(\l)$ from the pre-hydrodynamic mode only
\footnote{In practice, we truncate \Eq{psi-component} at $n=9$ so that $\sH_{{\rm RTA}}$ is reduced to a $10\times 10$ non-Hermitian matrix.
We have checked that the results shown are not sensitive to the truncation.}.
For comparison, we also determine $g(\tau)$ by solving the kinetic equation numerically.
Following Ref.~\cite{Heller:2018qvh},
we use the parametrization $\tColl \propto \e^{-\Delta/4}$.
Solutions to \Eq{eq:kinetic} with constant $\tColl$ ($\Delta=0$) and different initial conditions satisfying $\tau_I\ll \tColl$ are shown in dashed black in Fig.~\ref{fig:g}.
The resulting $g$ collapses to a common curve at times much earlier than $\tHydro$, which is the well-known ``attractor'' behavior of Bjorken-expanding RTA kinetic theory.  
Remarkably, 
$\tE_0(\l)$ is close to $g(\tau)$, 
indicating that the bulk evolution before the hydrodynamic regime is indeed dominated by the evolution of the pre-hydrodynamic mode.
In fact, the method developed in Refs.~\cite{rigolin2008beyond,APT} can be generalized to account systematically for the contributions from the excited modes $\phi_{n>0}$, as derived in \ref{sec:cond}.
Including leading-order contributions from the excited states to $g(\l)$, shown by the dashed red line, improves the agreement between the adiabatic result in \Fig{fig:g} and the RTA attractor.

Since the RTA attractor function $g(\tau)$ has already been obtained by many authors~\cite{Heller:2016rtz,Romatschke:2017vte,Heller:2018qvh},
what is our purpose of studying this function?
Our goal is to demonstrate that the main contribution to this function comes from the pre-hydrodynamic mode. 
We emphasize that the attractor behavior of $g(\l)$ alone does not tell us whether one mode or many modes are important for the subsequent evolution.
In the language of quantum mechanics, the attractor behavior only indicates that the system is in its instantaneous ground state around $\tAtt$.
The system remains in its instantaneous ground state here due to a qualitatively different reason, namely the suppression of transitions to the excited states.

To further demonstrate that the relative importance of contributions from excited modes are suppressed compared to those of the pre-hydrodynamic mode,
we show the fractional difference $\delta\equiv |\keyf-\tE_{0}|/\keyf$ as a function of $\l$ in Fig.~\ref{fig:diff}. 
Results for constant $\tColl$ ($\Delta=0$) and conformal $\tColl$ ($\Delta=1$) are shown in red and dashed blue, respectively. The energy density dependence of the relaxation time with $\Delta=1$ modifies the explicit $\tau$-dependence of $\lambda$.
The evolution is more adiabatic the smaller $\d$ is. 
$\d$ is small both when $\l\ll 1$ and $\l\gg 1$ 
and reaches a maximum of $0.045$ at intermediate $\l$.
This indicates that at least $95\%$ of the contribution to $\keyf$ between $\tAtt$ and $\tHydro$ is from the pre-hydrodynamic mode.  
We emphasize that $\d$ is small even when the Knudsen number $1/\l$ is large. 
To this order, $\delta$ is a proxy for the expansion parameter $\delta_A$ (see \ref{sec:cond}) that suppresses contributions from the excited states.

\section{Adiabaticity in the rapidly-expanding QGP
\label{sec:A-QGP}
}

Why does adiabaticity also apply to the violent expansion of the QGP in the early stages of the evolution?
In essence, ``adiabaticity'' only requires that the transition to excited states is suppressed. 
For example, consider a time-dependent Hamiltonian in quantum mechanics 
$H(t)=H_0+\tilde{\l}(t)H_1$, where $H_0, H_1$ are time-independent and $\tilde{\l}(t)$ is a monotonic function of time $t$.
The transition rate from the instantaneous ground state $|0,t\>$ to instantaneous excited states $|n,t\>$ is given by 
$\(\pd_{t}\log\tilde{\l}/\Delta E_n\)\,\<0,t|\tilde{\l}(t) H_1|n,t\>$~\cite{APT}. 
Therefore ``adiabaticity" can arise either due to the smallness of the rate of change of the Hamiltonian $\pd_{t}\log\tilde{\l}$ compared to the energy gap $\Delta E$ (slow-quench adiabaticity), or due to the time-dependent part of the Hamiltonian $\<0,t|\tilde{\l} H_1|n,t\>$ being small in amplitude (fast-quench adiabaticity), see Ref~\cite{APT} for examples of the applicability of adiabaticity to quantum phase transitions under fast quenches.

We have generalized the aforementioned quantum mechanical expression to a system described by \Eq{H-RTA} in \ref{sec:cond}.
While the slow-quench adiabaticity applies at late times as one might expect,
we also find that fast-quench adiabaticity applies at early times because $\l$ is small, 
see Eq.~\eqref{fast-Adi}. 
To see why this must be so on physical grounds, 
we recall that $\phi^{F}_0$ at very early times represents an angular distribution function $\Fe$ where typical values of $p_{z}$ are much smaller than those of $p_{\perp}$.
On the other hand, the excited states at early times have typical values of $p_z$ that are comparable to $p_\perp$.
A ``transition'' from the ground state to an excited state would therefore require either multiple scatterings or one rare hard scattering among gluons,
the probability of which is suppressed when $\tau\ll \tColl$.

Since our discussion above does not rely on the details of the collision integral, 
we expect that adiabaticity is a generic feature of both early- and late-time limits for the expanding weakly-coupled QGP.
In particular, consider the standard bottom-up thermalization scenario~\cite{Baier:2000sb}. 
Following the discussion above, 
we expect that adiabaticity applies during the stage $\tAtt\leq\tau\leq \a^{-5/2}_{s}\, Q^{-1}_{s}$ and $ \tau\geq \a_{s}^{-13/5}\, Q^{-1}_{s}$.
In the former stage, 
$\Fe$ represents the angular distribution of hard gluons (with typical energy $Q_{s}$) that rarely collide with one another.
In the later stage, 
$\Fe$ represents the angular distribution of soft gluons (with typical energy $T$) that
are already in thermal equilibrium.
Adiabaticity may break down during the transition stage $\a^{-5/2}\leq Q_{s}\tau\leq \a^{-13/5}$ when the numbers of both soft and hard gluons are changing rapidly, 
however this interval is parametrically narrow compared to other stages.

\section{Outlook}

While our analysis is based on a weakly-coupled kinetic description of the QGP, we anticipate that the concept of pre-hydrodynamic modes and the realization of \nameshort\ is relevant more broadly.
As we will show in upcoming work \cite{upcoming}, beyond Bjorken expansion in geometries with transverse expansion and spatial gradients the physics is richer. 
In this case not all slow modes at early times become hydrodynamic modes in the hydrodynamic limit, and not all hydrodynamic modes were necessarily slow modes before $\tHydro$, though the dominance of a reduced set of slow modes at early times remains.
It would also be interesting to explore \nameshort\ for the QGP at strong coupling~\cite{Bantilan:2018vjv}, and in table-top experiments~\cite{Brewer:2015ipa}. 
As a first step towards this exploration, 
it may be necessary to develop a more general method to identify pre-hydrodynamic modes from the pole structure of off-equilibrium correlation functions.

\sect{Acknowledgments}

We are grateful to 
J\"urgen~Berges, Jean-Paul Blaizot,
Min He, Ulrich~Heinz, Michal Heller,
Jiangyong Jia,
Wei Li,
Aleksi Kurkela,
Aleksas Mazeliauskas, Guilherme Milhano,
Mauricio Martinez Guerrero,
Krishna Rajagopal,
Paul Romatschke,
Chun Shen,
Misha~Stephanov, Viktor Svensson, and 
Urs Wiedemann
for helpful conversations.
This work was supported by the Office of Nuclear Physics of the U.S. Department of Energy within the framework of the Beam Energy Scan Theory (BEST) Topical Collaboration and under Contract  DE-SC0011090 (JB, YY), the Strategic Priority Research Program of Chinese Academy of Sciences, Grant No. XDB34000000 (YY) and the National Natural Science Foundation of China under Contract No. 11975079 (LY).

\appendix

\section{The condition for adiabaticity
\label{sec:cond}
}

In this section, we present a detailed discussion of the condition for the adiabaticity for the evolution of a vector $|\psi\>$ under
\begin{align}
\label{psi-right}
\pd_{y} |\psi(y)\>=- \sH(y)|\psi(y)\>\, . 
\end{align}
We shall assume that the non-Hermitian matrix $\sH(y)$ is non-degenerate, namely $\tE_{n}\neq \tE_{m}$ for $m\neq n$.
The discussion below is in parallel to the analogous study in quantum mechanics (see for example Ref.~\cite{APT}).
We define left eigenvectors $\<\phi_{n}|$ and right eigenvectors $|\phi_{n}\>$
\begin{align}
\,&\langle \phi_{n}(y) | \sH(y) = \tE_{n}(y) \langle \phi_{n}(y) |\,
\\
\label{LR-def}
\,&\sH(y)| \phi_{n}(y) \rangle=\tE_{n}(y)|\phi_{n}(y) \rangle \,
\end{align}
satisfying the orthogonality relations
\begin{eqnarray}
\langle \phi_{n}(y)|\phi_{m}(y)\rangle = \d_{mn}\, , 
\qquad
\sum_{m}|\phi_{m}(y)\rangle\langle \phi_{m}(y)| =I\, 
\end{eqnarray}
with $I$ denoting the identity matrix. 
Since $\sH$ is non-Hermitian, $| \psi\>$ and $\< \psi|$ are generally not related by Hermitian conjugation. 
We can expand general state vectors in the instantaneous eigenbases
\begin{align}
\label{psi-exp}
|\psi(y)\rangle = \sum_{n}\a_{n}(y)|\phi_{n}(y)\rangle\, ,
\qquad
\<\psi(y)| = \sum_{n}\tilde{\a}_{n}(y)\<\phi_{n}(y)|\, .
\end{align}
Substituting Eq.~\eqref{psi-exp} into Eq.~\eqref{psi-right}, 
we obtain 
\begin{align}
\label{a-fun}
\pd_{y}\a_{n}(y)= -\tE_{n}(y)\a_{n}(y)- \sum_{m\neq n}V_{nm}(y)\, \a_{m}(y)\, , 
\end{align}
where the evolution of the system leads to the contribution
\begin{align}
V_{mn}(y)\equiv  \langle \phi_{m}(y)|\pd_{y}|\phi_{n}(y)\rangle\, . 
\end{align}
In general, $V_{nn}(y)\neq 0$ and will enter into Eq.~\eqref{a-fun}. 
However, as discussed in Ref.~\cite{APT},
$V_{nn}$ can be chosen to be zero by rotating $|\phi_{n}(y)\>\to e^{-\int dy' V_{nn}(y')}|\phi_{n}(y)\>\ $.

Since the rates of change of the $\phi_{m>0}(y)$ modes are faster than that of the ground state mode $\phi_{0}(y)$, 
we expect that $|\a_{m>0}(y)/\a_{0}(y)|\ll 1$ on time scales longer than $1/|\Delta \tE_{m0}|$, where $\Delta \tE_{m0}=\tE_{m}-\tE_{0}$ is the gap. 
At this time scale, 
Eq.~\eqref{a-fun} for $\a_{m>0}$ can be written as
\begin{align}
\label{am}
\pd_{y}\a_{m}(y)=-\tE_{m}(y)\, \a_{m}(y)-V_{m0}(y)\, \a_{0}(y)\,, 
\qquad 
\textrm{($m>0$)}
\, .
\end{align}
Here we have dropped all contributions from $\a_{m>0}$ in the second term on the right hand side of Eq.~\eqref{a-fun}. 
The values of $\a_{m}(y)$ can then be estimated by finding the $\a_{m}(y)$ that makes the right hand side of Eq.~\eqref{am} vanish:
%
\begin{align}
\label{am-NLO}
\a_{m}(y)\approx -\frac{V_{m0}(y)}{\tE_{m}(y)} \a_{0}(y) \, . 
\end{align}
Eq.~\eqref{am-NLO} is the main result of this section. 
It takes into account transitions between the slow ($m=0$) mode and fast ($m>0$) modes due to the evolution of $\sH(y)$. 
It is also transparent from Eq.~\eqref{am-NLO} that the fast modes simply track the evolution of the slow mode. 
Substituting Eq.~\eqref{am-NLO} into Eq.~\eqref{psi-exp}, we obtain 
\begin{align}
\label{psi-adia}
|\psi(y)\> = \a_{0}(y)
\[ |\phi_{0}(y)\> -\sum_{m>0}\frac{V_{m0}(y)}{\tE_{m}(y)} |\phi_m(y) \>\]\, . 
\end{align}

The adiabatic approximation Eq.~\eqref{psi-adia} is valid when the adiabatic parameter $\delta_A$ is small,
\begin{align}
\label{d-V}
\delta_A \sim \left|\frac{V_{m0}(y)}{\tE_{m}(y)}\right|\ll 1.
\end{align}
To study the condition Eq.~\eqref{d-V},
it is convenient to express $V_{nm}$ in terms of $\sH$ by differentiating Eq.~\eqref{LR-def} with respect to $y$ and multiplying the result by $\<\phi_{m}(y)|$ to obtain
\begin{align}
\label{Vnm-H}
V_{nm}(y)= \frac{\<\phi_{n}(y)|\pd_{y} \sH(y)|\phi_{m}(y)\>}{\tE_{m}(y)-\tE_{n}(y)}\, ,
\qquad m\neq n\, . 
\end{align}
If we consider $\sH$ to be of the form Eq.~\eqref{H-RTA} then have from Eq.~\eqref{Vnm-H} that
\begin{align}
\label{d-V2}
\frac{V_{m0}(y)}{\tE_{m}(y)}=-\[ \frac{\pd_{y}\l}{\l\, \Delta \tE_{m0}(y)}\]\, \frac{\< \phi_{m}|\l \sH_{1}|\phi_{0}\>}{\tE_{m}(y)}\, . 
\end{align}
From Eq.~\eqref{d-V2}, 
we observe two sufficient conditions for $\delta_A \ll 1$.
The first is
\begin{align}
\label{slow-Adi}
\(\frac{\pd_{y}\l}{\l }\)\frac{1}{\Delta \tE_{m0}(y)}\ll 1\, , 
\end{align}
meaning the system will evolve adiabatically if the relative rate of change of $\l$ is small compared to the gap $|\Delta \tE_{m0}|$.
The second condition is 
\begin{align}
\label{fast-Adi}
 \frac{\< \phi_{m}|\l \sH_{1}|\phi_{0}\>}{\tE_{m}(y)}\ll 1\, ,
\end{align}
which corresponds to the ``fast-quench adiabaticity'' as mentioned in Sec.~\ref{sec:A-QGP} . 

With Eq.~\eqref{psi-adia} in hand,
it is possible to calculate quantities of interest to first order in $\delta_A$. 
In particular, we consider projecting onto the time-independent vector $\<0| \equiv \< \phi_0^H | = (1,0,0,\dots)$ that corresponds to the hydrodynamic mode in the hydrodynamic limit.
We multiply Eq.~\eqref{psi-right} by $\<0|$ to obtain
\begin{align}
g(y)= \frac{\< 0| \sH(y)|\psi\>}{\< 0|\psi\>}
\end{align}
where we have used Eq.~\eqref{psi-component} to obtain $\<0|\psi\>=\e$. 
Substituting Eq.~\eqref{psi-adia}
\begin{align}
\label{g-NLO}
g(y) \approx \tE_{0}(y)+\sum_{m>0} V_{m0}\(1-\frac{\tE_{0}}{\tE_{m}}\) \frac{\<0|\phi_{m}\>}{\<0|\phi_{0}\>}\, . 
\end{align}
In Fig.~\ref{fig:fig1}, 
we plot $g(y)$ computed from Eq.~\eqref{g-NLO}, 
which is almost identical to $g(y)$ obtained from solving Eq.~\eqref{eq:kinetic}.

\bibliographystyle{elsarticle-num}
\bibliography{refs}

\end{document}